# Design and Implementation Aspects of a novel Java P2P Simulator with GUI

V. Chrissikopoulos[a], G. Papaloukopoulos[b], E. Sakkopoulos[b], S. Sioutas[a,*]

[a] Informatics Dept., Ionian University, Corfu, Greece
[b] Computer Engineering & Informatics Department, University of Patras, Greece
{vchris,sioutas}@ionio.gr, {papalukg, sakkopul}@ceid.upatras.gr

**Abstract**

Peer-to-peer networks consist of thousands or millions of nodes that might join and leave arbitrarily. The evaluation of new protocols in real environments is many times practically impossible, especially at design and testing stages. The purpose of this paper is to describe the implementation aspects of a new Java based P2P simulator that has been developed to support scalability in the evaluation of such P2P dynamic environments. Evolving the functionality presented by previous solutions, we provide a friendly graphical user interface through which the high-level theoretic researcher/designer of a P2P system can easily construct an overlay with the desirable number of nodes and evaluate its operations using a number of key distributions. Furthermore, the simulator has built-in ability to produce statistics about the distributed structure. Emphasis was given to the parametrical configuration of the simulator. As a result the developed tool can be utilized in the simulation and evaluation procedures of a variety of different protocols, with only few changes in the Java code.

## 1. Introduction and Motivation

In this paper we describe the design and implementation aspects of a novel P2P simulator. The developed simulator is based on a Message Passing environment where the peers are represented by Java threads and communicate with each other by sending and receiving messages. Three central levels can be distinguished in the developed tool. The first level is the system's kernel where the peers, the network environment and the messages have been implemented. The second level is the user interface of the simulator which facilitates the P2P operations of the network under simulation (node joins, updates, leaves etc). And finally, the third level is the graphical user interface of our system that facilitate the evaluation and statistical analysis.

There are a number of simulators presented in the P2P literature. An extensive and analytical peer to peer simulators survey has been presented recently in [7]. A large number of papers (~70) have been included in the survey, which stated which simulator they used. The majority of these papers (62%) used a specially created simulator for their own case study evaluation. Some of these simulators might possibly be the same, reused within research groups. However, even taking this into account, the number of custom-made simulators far outnumbers the use of known simulators. This is not an ideal state of affairs, both in terms of duplication of effort and for ease of comparison and replication of results. The authors of the survey believe that "the poor state of existing P2P simulators is the reason that much published research makes use of custom built simulators".

Overall Neicken et al [7] have found that a key drawback of the simulators is that they have no systematic and user friendly mechanisms to allow a user to gather statistics of a simulation run that is what leads to different evaluation-simulation tools for new P2P environments.

To overcome such difficulties we propose a novel P2P simulator tool, which includes built-in features to facilitate the logging and production of systematic statistical results for the algorithms' performance as well as the load balance of the distributed structure. Furthermore its friendly graphical user interface allows non-programmers - high-level researchers/designers of P2P systems to easily construct their overlay with the desirable number of nodes and evaluate its operations using a number of key distributions. Our prototype is publicly available and free to download [8]. We evaluated our prototype simulator using the NBDT (Nested Balanced Distributed Tree) [1], a fault tolerant discovery-search infrastructure for P2P Web Service discovery.

---

* Contact author: Spyros Sioutas sioutas@ionio.gr; authors appear alphabetically

The rest of this paper is organized as follows. Section 2 provides a short discussion of related popular P2P simulators. A short overview of the NBDT Network is given in section 3. Section 4 introduces the systems architecture. Section 5 analyses the peer/node structure. Section 6 and 7 describe the user interfaces. In section 8, experimental results are illustrated. Finally, section 9 concludes the paper.

## 2. Related Work on P2P simulators

In the comparative P2P simulator survey of Neicken et al [7], there is a number of tools discussed. For independency, popular solutions are shortly discussed here.

PeerSim [4] has been developed with extreme scalability and support for dynamicity in mind. It is composed of two simulation engines, a simplified (cycle-based) one and event driven one. PeerSim is written entirely in Java. Unfortunately it completely lacks of graphical interface support as well as visualization interfaces, making it difficult to be used, especially by high level algorithmic/designers P2P researchers.

p2psim [5] is a free, multi-threaded, discrete event simulator to evaluate, investigate, and explore peer-to-peer (p2p) protocols. p2psim is part of the IRIS project. p2psim supports several peer-to-peer protocols, making comparisons between different protocols convenient. p2psim is developed using C++ while most simulators are Java based. This makes its API difficult to support and to extend further. Unfortunately, it has scalability limitations in number of nodes (<3000).

OverSim [6] is an open-source overlay network simulation framework for the OMNeT++/OMNEST simulation environmen. OMNeT++ is open-architecture simulation environment with strong GUI support and an embeddable simulation kernel. Its primary application area is the simulation in the field of internet simulations (IP, IPv6, MPLS, etc), mobility and ad-hoc simulations. OverSim utilizes the GUI support of OMNeT++ to provide a node visualization mechanism. However such visualization mechanisms are only useful in network simulations with small numbers of nodes. Evaluation of large scale simulation network is not supported by the GUI but using external, post processing Perl based scripts. Furthermore the OMNeT++ infrastructure is based on detailed modelling of lower network layers that are of little interest to P2P researchers and make the overlay networks evaluation more complex.

## 3. Overview of the NBDT network

The structure that we examined was built by repeating the same kind of BDT tree-structure (Balanced Distributed Tree) in each group of nodes having the same ancestor, and doing this recursively. This structure may be imposed through another set of pointers. The innermost level of nesting will be characterized by having a tree-structure, in which no more than two nodes share the same direct ancestor. Figure 1 illustrates a simple example (for the sake of clarity we have omitted from the picture the links between nodes with the same ancestor). Thus, multiple independent tree structures are imposed on the collection of nodes inserted. Each element inserted contains pointers to its representatives in each of the trees it belongs to.

Let $\sigma_k$ an initial given $\mu(\cdot)$ sequence of *w-bit* keys belonging in universe *K=[0,2$^w$-1 ]*, where $\mu(\cdot)$ an unknown density. At initialization step we choose as peer representatives the *1$^{st}$ key*, the *lnK$^{st}$ key*, the *2lnK$^{st}$ key* and so on, meaning that each node with label i (1≤i≤N) stores ordered keys that belong in range [*(i-1)lnK,..ilnK-1*], where *N=K/lnK* the number of peers. Note that during update operations; it is not at all obvious how to bound the load of the *N* peers, since new w′-bit keys with w′>w may be appeared in the system and *K* must exceed. For this purpose we will model the insertions/deletions as the combinatorial game of bins and balls presented in [2]: Modelling the insertions/deletions of keys in this way, the load of each peer becomes $\Theta(poly\log N)$ in expected case with high probability. Obviously, peers' representatives early described have also been chosen according to this game. We also assume that each key is distinct and as a result the probability of collisions is zero. Each key is stored at most in *O(loglogN)* levels. We also equip each peer with the table *LSI* (Left Spine Index). This table stores pointers to the peers of the left-most spine (for example in figure 1 the peers 1, 2, 4 and 8 are pointed by the LSI table of peer 5) and as a consequence its maximum length is *O(loglogN)*. Furthermore, each peer of the left-most spine is equipped with the table *CI* (Collection Index). *CI* stores pointers to the collections of peers presented at the same level (see in figure 1 the *CI* table of peer 8). Peers having same father belong to the same collection. For example in the figure 1, peers 8, 9, 10, and 11 constitute a collection of peers. It's obvious that the maximum length of *CI* table is $O(\sqrt{N})$.

For example in figure 1 we are located at (green) node 5 and we are looking for a key k∈[13lnn, 14lnn-1]. In other words we are looking for (green) node 14. As shown in [1], the whole searching process requires: $T = \sum_{k=1}^{\log \log N} O(1) = O(\log \log N)$ hops or lookup messages and that is also validated using the proposed simulator.

When we want to insert/delete a key/node from the structure, we initially search for the node that is responsible for it (using a number of O(*loglogN*) hops in worst-case) and then we simply insert/delete it from the appropriate node.

If new w′-bit keys, with w′>w, request to be inserted into the system, then we have to insert new peers on the network infrastructure and as a result we have to re-organize the whole p2p structure. In practice, such an expensive re-organization is very sparse. The new peers of NBDT are inserted at the end of the whole infrastructure consuming O(1) hops in worst-case. In particular, when a node receives a joining node request it has to forward the join request to the last node. The last node of NBDT infrastructure can be found in O(1) hops in worst-case by using the appropriate LSI and CI indexes.

If the load of some peer becomes zero, we mark as deleted the aforementioned peer. If the number of marked peers is not constant any more then we have to re-organize the whole p2p structure. Based on the basic theorem of [2], if we generate the keys according to smooth distributions, which is a superset of regular, normal, uniform as well as of real world skew distributions like zipfian, binomial or power law (for details see [3]), we can assure with high probability that the load of each peer never exceeds *polylogn* size and never becomes zero. The latter means that with high probability split or delete operations will never occur. In other words, the re-organization of the whole P2P structure with high probability will never occur.

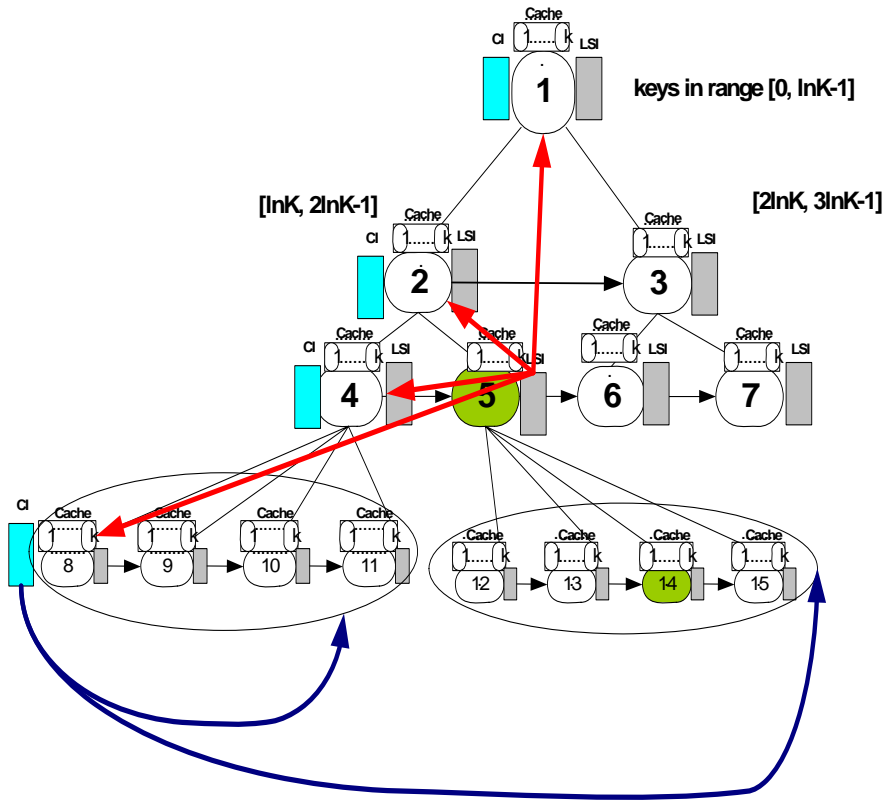

**Figure1.** The NBDT peer-to-peer system

## 4. Architecture: The message passing environment

The basic architecture of the P2P simulator is based on the message passing environment which is discussed in this section. The peers exchange messages in order to build the overlay network and to carry out the search, insert and delete operations of a key. These are two different types of messages. In the first type the message has information about the LSI and CI tables, about the parent of the inserted node, its sibling node and the total number of nodes in the overlay network. On the other hand, the message which is being send during a search, insert or delete operation only contains the key for these operations. The information about the transmitter's and the receiver's address are included in both types of messages.

This form of message is like the IP's format message. Having this in mind, the message that will be transmitted in the network is constructed. This message has the transmitter's node Id, the receiver's node Id, the type of the message and final, the data.

For this purpose we have implemented a class Message and a class Data. The class Message has four private fields which hold the id of the transmitter and the receiver, the type of the message and a pointer to the Data object. Also, this class has static final variables which declare the type of the message (join, insert, delete, etc.).

Next, the Network environment will be described. In our implementation, we make the assumptions that the messages don't get lost and no delay exists expect for the delay that take place between the time of transmitting and the time a specific node will take notice about the arrival of a new message. For the beginning, we don't concern about the network failures and the delays. Later we will discuss how we can add these features in our network but firstly, the simple network model will be described. To simulate the network's behavior we have implement the class Network which consist of a buffer, a counter and a file descriptor. The buffer is implemented with the Vector type of Java and stores objects of type Message.

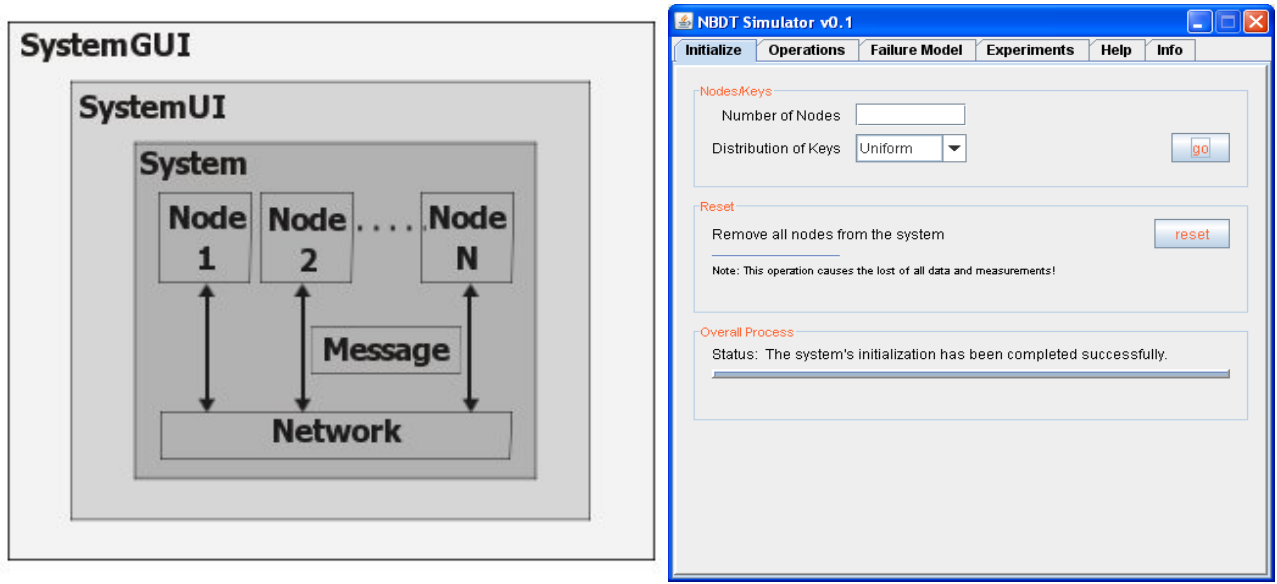

**Figure 2 and 3.** The Simulator architecture.
Simulator's tab "Initialize"

The counter is used for counting the total number of messages which delivered at an operation time as well as the file descriptor stores the appropriate messages about the process is being executed. This class also implements several methods such as: sendMessage(), recvMessage(), broadcast() and msgForNodeId(). When a node wants to send a message it calls the method sendMessage which takes one argument, the message. This method which is synchronized attends to store the message into the buffer and increase the counter by one. If the type of the message is "*search*", "*insert*" or "*delete*", then the algorithm writes to the file a record with the following format: i.e. "*Search message for node 3 to node 45*." This log file is used by the graphical user interface to show how an operation is incrementally completed. On the other hand, each node observes consecutively the network in order to verify if any message has been arrived. This operation can be done with Network's method msgForNodeId. If there is a message for this node the method returns the index where the message will be stored into the buffer, otherwise it returns -1. In the first case, the node can receive the message by calling the method recvMessage. As we can see the method sendMessage is non-blocking, while the method recvMessage is blocking. In this way, we are simulating the functions *send* and *receive* of POSIX C. The broadcast() method is used during the construction of the overlay for sending the same message over a range of nodes. This method which is also synchronized is a wrapper function of sendMessage and is used only at the time of the network construction, since no broadcast is needed during a search operation. In the next section we will proceed on the analysis of the peers' functionality and their structure.

## 5. The Peer Nodes

In this section we analyze the structure of the NBDT's nodes and their functionality. Some peer to peer simulators [4][5] use only limited threads to simulate the network in a sense of the producer/consumer model. With this synchronous approach one thread parses all the nodes of the network and produces the messages while the other thread parses the nodes and consumes the messages that have been generated by the first thread. In our simulator we have follow a different approach where each node is a thread and each of them is executed independently. Although this model is realistic enough, it also requires huge memory consumption and as a consequence we can simulate less and less nodes. Additionally, we have to face the scheduling problem since the Java's runtime environment does not provide scheduling processes. While each node is represented by a thread which runs ceaselessly and since there are critical segments, another problem we have to avoid is the thread competition. There are two critical segments: the incoming nodes must be inserted sequentially and only one node must operate on the network's buffer at a time.

```
public  void class Node extends Thread{
public Node () {
//allocate the memory for the vectors }
public void run() {
        while(!stopThisThread)
                eventHandler();}
private void eventHandler() {
        if (net.msgForNodeId(myID)) {
                msg = net.recvMessage();
                resolveMessage(msg);
        }}
private void resolveMessage(Message msg) {
        if (msg.getType() == JOIN)
             forwardJoinMessage(msg);
        else if (msg.getType() == SEARCH)
             forwardSearchMessage(msg); }
private void forwardJoinMessage(msg) {
        if (myID != lastLSNode)
//send message  to last left spine node
        else if (myID == lastLSNode)
//send message to last node of the last collection
        else if (sibling == empty)
//append new node and send back to the incoming
//node  the LSI table, its ID, etc..}
private void forwardSearchMessage(msg) {
        nodeID = findKeyRange(msg.key);
     if (myID == nodeID)
//search locally for the desired key and send back
//the response.
```

```
      else {
         if (LSNodeId != myID && !isLSNodeIdVisited)
//Forward the search message to the LS node.
         else if (myID == LSNode)
//send the search message to the representative
//of the right collection of this level
         else {
//This is the root node of the right collection.
//Reapeat the same procedure at the next nested
//level tree.}}}
```

**Table 1: Pseudo-Code for the class *Node***

To simulate the NBDT's nodes we have constructed the class Node which extends the Java's Thread class. This class holds the tables which are necessary for constructing the overlay and provides the methods for joining an incoming node, deleting an existed node, searching and updating a key. The threads communicate via the network exchanging messages as described in the previous section.

If a new node wants to join, the network has to send a join message to one of its introducer nodes. The introducer node listens to the request and forwards the message to the last node. This node is responsible for sending to the incoming node its routing table, the id of the incoming node and all the rest necessary information. Moreover, the last node updates its pointers to point to the new node. The procedure is repeated recursively for each nested level until the node is inserted to all possible levels. Table 1 presents a pseudo code for the class Node having focused on the join and search operations.

The crucial point here is that the incoming nodes must be inserted sequentially in order to construct properly the tree. Assume that two nodes want to join the network at the same time, then the introducer will forward their request to the last node of the tree.

**6. System User Interface**

To be able to deal with our system, we have implemented a class SystemUI which is the interface of our system. This class provides the methods for the initialization of our system, the functions for the search, insert and delete operations of the NBDT and methods which provide general information about the status of our structure as the total number of keys and nodes, the load balance and the range of the keys. This is the main class which also starts up the graphical user interface which we describe above.

More specifically, the SystemUI class initializes an overlay with only three nodes which are the introducer nodes for those nodes that want to join the overlay. These nodes never fail, so that other nodes can join the system. Then, we can insert a desired number of nodes in the system and store some keys according to a given distribution (uniform, normal, beta and pow-law). Furthermore, we initialize the Network as well as a vector that will store the new nodes.

**7. Graphical User Interface**

Someone can easily built a new P2P network and carry out experiments to evaluate the efficiency of this protocol through a graphical user interface. GUI consists of a window with six tabs, each of them is separated with panels to distinguish the different available operations. The graphical user interface has been implemented with NetBeans 5.5 software. In the following, we describe the functionalities of each tab.

In the first tab the user can set the number of nodes which will constitute the overlay and select the key distribution over these nodes. The available distributions are: uniform, normal, beta, and pow-law. After the user has set these two fields then the system's initialization can begin. In the same tab there is a progress bar so the user can obtain the overall process due to the fact that this process may take several minutes. Also there is a button, which resets the system without the need of closing and reopening the simulator if we want to carry out several experiments with different number of nodes and key distribution.

In the second tab – operations - the user can see the current number of nodes into the system, the number of keys that have been stored over the nodes and the range of keys that we can store in the overlay. The other three panels provide the ability to search, insert and update a key starting the procedure from any node in the network. While one of these operations is being executed, appropriate messages are appearing at the bottom of this tab.

In the forth tab the user can prosecute experiments to evaluate the efficiency of the NBDT's algorithms. There are three panels one for each operation where the user sets the number of the experiments and selects the distribution according to the keys will be picked up for the experiments. After the termination of the experiments the user can see and save in png format the chart that has been generated. Furthermore, this tab generates a chart with the key distribution over the nodes (load balance). The charts are generated by the free distributed software JCharts and the distributions generated by the cern package, both of them implemented with Java. The rest two tabs provide information about the usage and design of this software.

## 8. Simulation and Results

In this section we evaluate the NBDT protocol by the simulator described in previous sections. More specifically we evaluate the search path length and the load balancing performance for the four key distributions: uniform, normal, beta and pow-law.

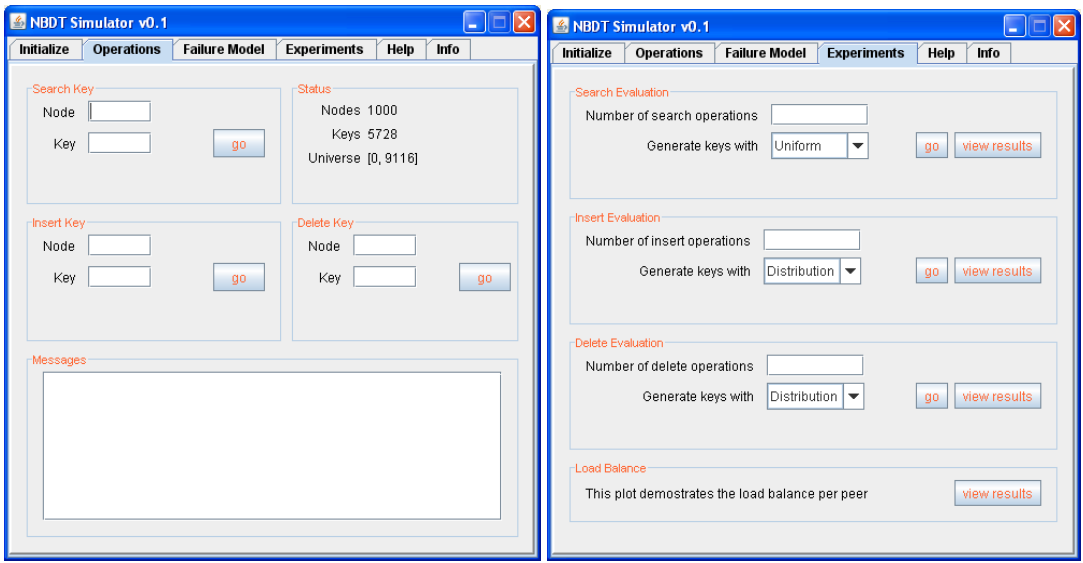

**Figure 4 and 5**. The tab "Operations" and "Experiments".

In order to understand in practice the routing performance of this protocol, we simulated a network with N=[1000, 5000] nodes, storing K=[5728, 22572] keys in all. We conducted a separate experiment of each value. Each node in an experiment picked a random set of keys to query from the system, and we measured the path length required to resolve each query. For the experiments we considered synthetic data sets. Their generation was based on different distributions to choose from. A sample of the network's lookup efficiency is depicted in figure 6.

To evaluate the load balancing of our structure we performed several experiments. Initially the keys for the nodes were picked up according to a chosen (configurable) distribution, then we performed 2500 updates where the inserted/deleted keys draw the same (or configurable other) distributions. Figure 7 shows that NBDT has smoothly distributed load per peer, and this holds as we increase the number of nodes and keys in the network.

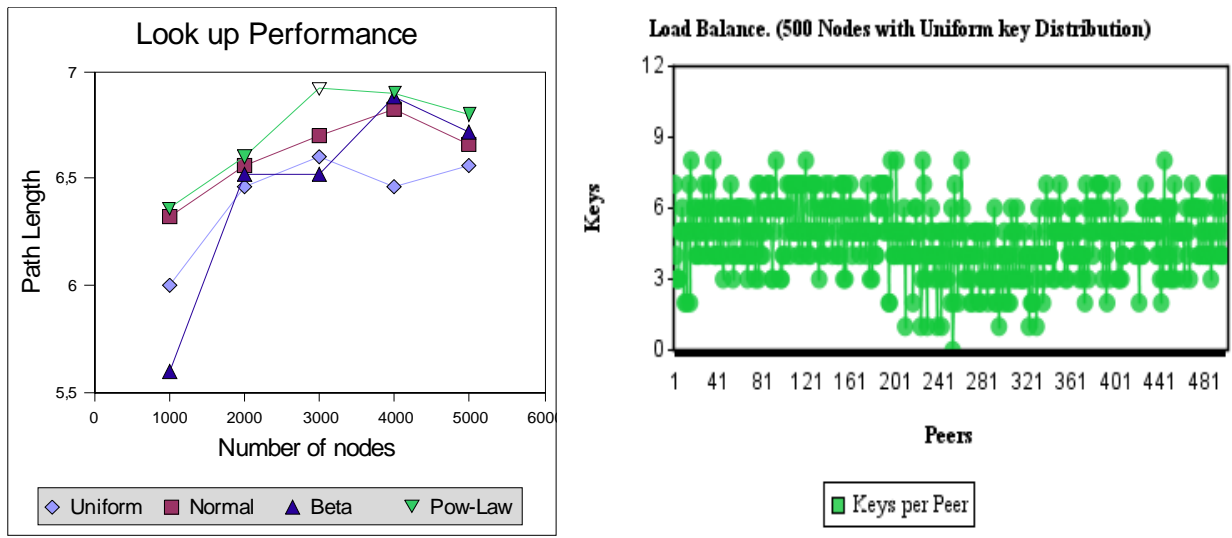

**Figure 6 and 7**: Look up performance graph, Load balance after 2500 updates with pow-law distribution.

## 9. Conclusions

In this paper we described the design and implementation aspects of a novel Java P2P simulator. The key features of the simulator is its extensibility, the user interface and the statistics support. The methodology presented can be easily extended to simulate real applications that use the different P2P protocol. The latter requires a few changes only, in the main class which implements the functionalities of the protocol. The prototype was evaluated upon the NBDT [1] solution. These changes are bounded in the message passing environment where the methods that simulate our network have to be replaced by the Java's network programming methods. Furthermore, we have developed a friendly graphical user interface, which provide all the main operations of a P2P protocol and has the ability to export directly charts with the algorithms' performance as well as the load balance per peer. Our aim is provide this friendly GUI Java based simulator as a basic tool of a framework towards the standardization of P2P simulations that will user friendly facilitate different protocol evaluation and comparison. Future work includes the support more simulation hosts.